\documentclass[a4paper]{VisionStyle}  
\usepackage{epsfig}  
  
\begin{document} 
 
\title{An XMM-Newton Survey of Ultra-Luminous Infrared Galaxies} 
 
\author{V.\,Braito\inst{1,2}, A.\,Franceschini\inst{2}, R.\,Della Ceca\inst{3},   
L.\,Bassani\inst{4}, M.\,Cappi\inst{4}, P.\,Malaguti\inst{4}, 
G.G.C.\,Palumbo\inst{5}, M.\,Persic\inst{6}, G.\,Risaliti\inst{7}, 
 M.\,Salvati\inst{8} and P.\,Severgnini\inst{3}} 
\institute{ 
  Osservatorio Astronomico di Padova, Italy 
\and  
  Dipartimento di Astronomia, Universit\`a di  Padova, Italy 
\and  
Osservatorio Astronomico di Brera, Milano, Italy 
\and 
TESRE/CNR, Bologna, Italy 
\and 
Dipartimento di Astronomia, Universit\`a di Bologna,  Italy 
\and 
Osservatorio Astronomico di Trieste,  Italy 
\and 
Dipartimento di Astronomia e Scienza dello Spazio, Universit\`a di Firenze, Italy 
\and 
Osservatorio Astrofisico di Arcetri, Firenze, Italy. 
} 
\maketitle  
 
\begin{abstract} 
 
We present preliminary results of XMM-Newton observations of 5 Ultra-luminous Infrared Galaxies
(ULIRGs), part of a mini-survey program dedicated to 10 ULIRGs selected from the bright IRAS
sample.  For 3 of them (IRAS 20551-4250, IRAS 19254-7245 and Mkn 231) we find strong  
evidence for the presence of a hidden  AGN, while for two others (IRAS 20110-4156 IRAS 22491-1808)
the S/N ratio of the data does not allow us to be conclusive.  
In particular, we have detected a strong Fe-K line in the X-ray spectra of  
IRAS19254-7245, with an equivalent width ($\sim 2$  keV) 
suggestive that most of the energy source in this object is due to a deeply buried AGN.
\end{abstract}

\section{Introduction} 
 
The nature of the energy source in Ultra-luminous Infrared Galaxies (ULIRGs),  
sources with L$_{IR}> 10^{12}$L$_\odot$, is still a debated issue.  
Although it is widely accepted that both starburst and/or AGN activity may be responsible 
for the observed luminosity, their relative contribution is still unconstrained. 
In some cases, even the presence of an AGN is unclear.   
 
Hard X-ray (E$>$2 keV) studies offer a fundamental tool to investigate the presence of hidden AGNs
inside ULIRGs and to obtain quantitative estimates of their contribution to the bolometric
luminosity.  
Indeed some of the  ULIRGs, classified as  pure starburst based on infrared and optical
spectroscopy (i.e. NGC6240, \cite{vbraito-C2:Iwasawa1999}),   
show spectral properties typical of obscured AGNs, when observed in the hard X-rays. 
 
Previous X-ray studies of ULIRGs with ROSAT, ASCA and BeppoSAX  
have shown that ULIRGs are often faint X-ray sources (e.g. \cite{vbraito-C2:Risaliti2000}).   
Combined with their high luminosity in the infrared bands, this  
could suggest the presence of an obscured AGN.  
Since ULIRGs are the major contributors to the infrared cosmological background, clarifying the
issue of how much of their energy output comes from obscured AGNs would have 
important implications on both models for the origin of the IR background, and
the IR to X-ray background connection. 

To shed light on these problems we are carrying out a mini-survey with XMM-Newton of 10 ULIRGs
for which high quality mid-IR and optical spectroscopic data are available.  
We present here preliminary results on 5 sources, two of which were not 
detected before in X-rays.

\section{The ULIRGs SAMPLE} 
\label{vbraito-E1_sec:XMM} 
The 5 ULIRGs discussed here have been selected from the sample of Genzel el al. (1998)
which is complete to 5.4 Jy at 60 $\mu$m and includes sources with $L\geq 10^{12}$ L$_{\sun}$
in the 12--100 $\mu$m band. The FIR selection makes the sample unbiased with respect  
to absorption, and questions like the relative importance of starburst versus AGN activity
can be discussed on a sound physical and statistical ground. In addition, high quality  
infrared and optical spectroscopic data are also available for all the  
sources in the sample (\cite{vbraito-C2:Lutz1999}, \cite{vbraito-C2:Veilleux1999}).  
The infrared luminosity and redshift of these 10 objects,
as well as the XMM-Newton exposure time  
of the 5 objects discussed here, are reported in Table~\ref{vbraito-C2_tab:tab1}. 
Optical and infrared classifications of the  5 ULIRGs presented here are summarized in  Table~\ref{vbraito-C2_tab:tab2}.
All ten objects are being observed with XMM-Newton with $\sim$20 ksec exposure each.
The main goal of these  observations was to detect the hard X-ray continuum and to investigate the presence of features indicative of AGN activity such as Fe-K lines. 
The original Genzel sample is composed of 15 ULIRGs: 4 of the remaining 5 objects
are being observed anyway with XMM-Newton by different PI's (data will become public
after 1-year proprietary period).
For the last object (IRAS 23060+0505) adequate ASCA data are already public. 
Our aim will also be to combine our X-ray with other available data (XMM-Newton,
{\it Chandra}, ASCA), as well as with information at other wavelengths.

\section{XMM EPIC: Data Preparation and Analysis} 
\label{fauthor-E1_sec:an} 
 
The XMM-Newton observations presented here  
have been performed between March 2001 and December 2001 with    
the EPIC cameras operating in full-frame mode. 
Data have been processed using the Science Analysis Software (SAS version 5.2); 
the latest known calibration files and response matrices released by the EPIC team have been used.  
  
Event files released from the SAS standard pipeline have 
been filtered from high-background time intervals and only events  
corresponding to pattern 0-12 for MOS and pattern 0 for PN have been used  
in the scientific analysis. 
The screening process from high background time intervals yielded net exposures between 16.3 ksec
and 21.8 ksec (net exposures for PN camera are reported in  Table~\ref{vbraito-C2_tab:tab1}).  
 
All the 5 ULIRGs have been detected in the MOS1, MOS2 and PN detectors  
with signal to noise ratios greater than 3.  
Except for Mkn 231, which appears to be slightly extended,  
source spectra were extracted from circular regions of $\sim 15$ arcsec radius.  
Background spectra have been  
extracted from source-free circular regions of $\sim 1 $ arcmin radius   
close to the target.
Mkn 231 appears clearly extended ($\sim$ 40 arcsec) in
the XMM-Newton images in particular in the soft X-ray band (E$<$ 2 keV).
In order to be less affected by the galaxy host emission we have chosen
to extract the spectra from  what appears to be  the hard X-ray core of Mkn 231 (radius 15 arcsec).
 In order to improve the statistic, MOS1 and MOS2 data have been combined. 
Both MOS and PN spectra were then rebinned to have a S/N $\geq 2$ in each energy channel.  
The combined MOS data have been  fitted simultaneously with PN data,  
keeping free the relative normalizations. 

\begin{table}[bht] 
  \caption{Summary of the XMM-Newton  accepted targets } 
  \label{vbraito-C2_tab:tab1} 
  \begin{center} 
    \leavevmode 
    \footnotesize 
    \begin{tabular}[h]{lcccl} 
      \hline \\[-5pt] 
    Name    &z&Log L$_{IR}$      &  EXP\\[+5pt] 
      \hline \\[-5pt] 
 
  &  &                      ($10^{11}$ L$\odot$)                    & ksec \\  
\hline 
IRAS 12112+0305$^{a}$&0.072&17&/\\ 
MKN231$^{b}$&0.042&30&16.30\\ 
IRAS 14348-1447$^{a}$&0.082&18&/\\ 
IRAS 15250+3609$^{c}$&0.055&8.8&/\\ 
IRAS 17208-0014$^{c}$&0.043&21&/\\ 
IRAS 19254-7245&0.062 &10 &18.39\\ 
IRAS 20100-4156&0.129   &33&18.09\\ 
IRAS 20551-4250& 0.043&  9.5&16.31 \\  
IRAS 22491-1808& 0.078&12&21.77\\ 
IRAS 23128-5919$^{c}$ &0.044&9.1&/\\

 \hline \\ 
      \end{tabular}
 \end{center} 
  \textit{Note. a: source observed, data to be processed; b: very preliminary results; c: source to be observed.}

\end{table}

\section{Preliminary Results} 
\label{vbraito-E1_sec:SPECTRA}  
 
\subsection{IRAS 22491-1808 and IRAS 20100-4156} 
 
IRAS 22491-1808 and IRAS 20100-4156 have been detected with a low significance
which prevents us to perform a detailed spectral analysis. For these two sources
we tried only two simple models: a thermal model and an absorbed power-law one. 
For both sources a single thermal or a single power-law model are both rejected from the data.
The X-ray spectra of these two objects can be described with thermal model
(kT=$0.86_{-0.08}^{+0.19}$ keV for IRAS 22491-1808 and  
kT=$0.67^{+0.37}_{-0.30}$ keV for IRAS 20100-4156), but evidence of an excess  
in the observed spectra at E $>$ 2 keV is present. 
In order to account for this hard emission an absorbed  power-law component  
has been added to the thermal model. 
Fixing $\Gamma=$ 1.7 we obtain $N_H \sim 10^{20}$cm$^{-2}$ for IRAS 22491-1808 and  
$N_H \sim 10^{22}$cm$^{-2}$ for IRAS 20100-4156. The soft X-ray ($0.5-2$ keV)
luminosities, corrected for absorption,  are $\sim 2\times 10^{41}$erg s$^{-1}$ (IRAS 22491-1808) and
$\sim 5\times 10^{41}$erg s$^{-1}$ (IRAS 20100-4156).
The intrinsic hard ($2-10$ keV)
X-ray luminosity is  $\sim 10^{41}$erg s$^{-1}$ for IRAS 22491-1808 
and $\sim 5\times 10^{41}$erg s$^{-1}$ for IRAS 20100-4156. 
 
\subsection{IRAS 20551-4250} 
 
\begin{figure}[!t] 
 \begin{center} 
   \epsfig{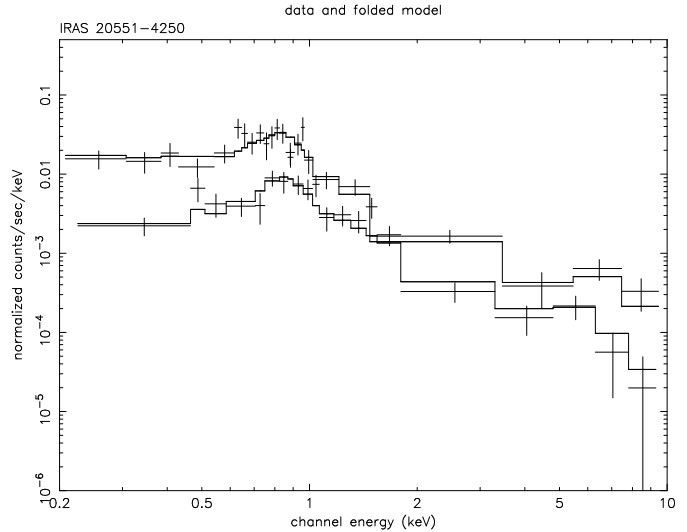} 
 
\caption{\textit{IRAS 20551-4250: MOS, PN data and folded model.}} 
\label{vbraito-C2_fig:fig1} 
\end{center} 
\end{figure}

This source  was previously observed with ASCA (\cite{vbraito-C2:Misaki1999}) but the  
S/N ratio was very low. 
Thanks to the XMM-Newton throughput this source is now detected  
with an adequate  S/N to perform a detailed spectral analysis.  
A good fit  (see Figure~\ref{vbraito-C2_fig:fig1}) is obtained with a two components model: a thermal model (kT=0.68$^{+0.12}_{-0.07}$ keV) plus the so called ``leaky-absorber"  
continua (this latter component is composed by an absorbed plus  
a non-absorbed power law model having the same photon index).  
We found that   
a good fit can be obtained with  $\Gamma=1.78^{+0.23}_{-0.24}$,   
a partial covering  factor equal to $\sim 93\%$ and N$_H=8.6\times10^{23}$cm$^{-2}$. 
The Iron line at 6.4 keV is not detected but the upper limit  
on its equivalent width ($\sim 1$ keV) is consistent with the  
$N_H$ value obtained from the fit. 
The intrinsic hard X-ray luminosity, corrected for absorption, of this object is $\sim 2\times 10^{42}$
erg s$^{-1}$  the soft X-ray luminosity is
$\sim 10^{42}$ erg s$^{-1}$.

\subsection{IRAS19254-7245} 
 
From the  analysis of the ASCA data (\cite{vbraito-C2:Imanishi1999};  \cite{vbraito-C2:Pappa2000}) two acceptable fits   
were proposed for this object: 
a) a non-absorbed power law model having a flat photon index ($\Gamma\sim1$),
and    b) an absorbed power-law
model with $\Gamma$ fixed to 1.7 or 1.9 and N$_H\sim 10^{22}$ cm$^{-2}$. 
No Iron line was clearly detected. 
 
\begin{figure}[!t] 
\begin{center} 
\epsfig{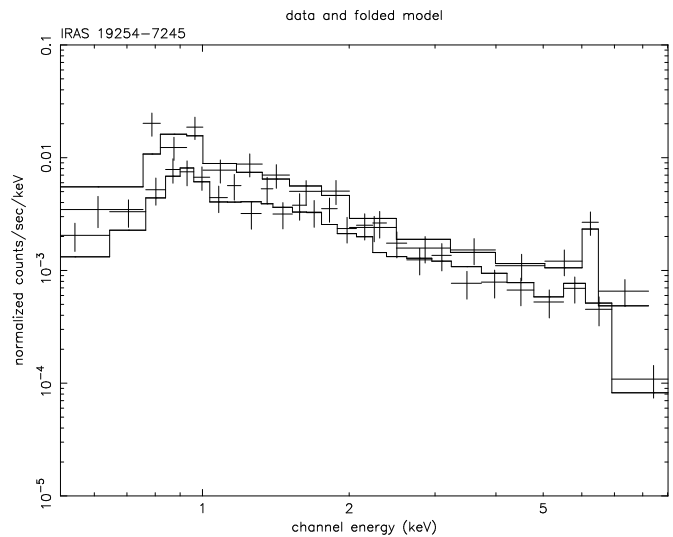} 
\caption{\textit{IRAS 19254-7245: MOS, PN data and folded model.}} 
\label{vbraito-C2_fig:fig2} 
\end{center} 
\end{figure}

A single unabsorbed power law model is not a good description of the XMM-Newton data
because of large discrepancies in the low energy domain. 
We have then added a thermal component obtaining a temperature kT=$0.98$ keV  
for the soft component and $\Gamma=1.20^{+0.25}_{-0.30}$ (N$_H=3\times10^{21}$cm$^{-2}$) 
for the hard component (see Figure~\ref{vbraito-C2_fig:fig2}).  The soft X-ray luminosity is
$\sim2\times 10^{42}$ erg s$^{-1}$. 
We also have evidence for an Iron line at 6.4 keV, whose high equivalent width
($1.6$ keV), together with the flat photon index, indicates that this
object could be ``Compton thick", with the detected hard X-ray emission due
to a pure reflected component.          The intrinsic luminosity of the latter could be
higher than the derived value of $\sim 10^{42}$ erg s$^{-1}$.

\subsection{MKN 231} 
 
Among the local ULIRGs, Mkn231 is the most luminous object ( \cite{vbraito-C2:Soifer1984}) and is one of the best studied at all wavelengths.
Although it has been observed with many X-ray observatories (ROSAT, ASCA and more recently  
with \textit{Chandra}), its X-ray properties remain still puzzling.  

The AGN and starburst activities were clearly evident
from ROSAT and ASCA data  (\cite{vbraito-C2:Iwasawa1999}; \cite{vbraito-C2:Turner1999}).
However the flatness of the X-ray spectra at energies above 2 keV and the lack of
any strong Fe line  (\cite{vbraito-C2:Maloney2000}) was unusual;

these results have been confirmed by recent Chandra observations
($\Gamma=1.3$, EW$<$188 eV; \cite{vbraito-C2:Gallagher2001}).

We have done only a preliminary analysis of the XMM-Newton data on Mkn 231
(the field containing the source is shown in Figure~\ref{vbraito-C2_fig:fig3}).
Our results confirm the presence of a very flat spectrum ($\Gamma=0.9$) and  only a marginally detected, moderately intense, Iron line (EW$\sim$200 eV).  
The intrinsic luminosity of the hard component is $\sim 3\times10^{42}$ erg s$^{-1}$. 
A more complete analysis of Mkn 231 will be presented in a forthcoming paper.

\begin{table}[t] 
  \caption{Observed ULIRGs Classification} 
  \label{vbraito-C2_tab:tab2} 
  \begin{center} 
    \leavevmode 
    \footnotesize

\begin{tabular}{lccc} 
\hline 
\bf{Name }        &\bf{Optical} & \bf{Mid-IR}& \bf{X-ray} \\     
 (1) &(2)     &(3)                     &(4)       \\  
\hline 
 
IRAS 20100-4156&HII&SB&AGN??\\ 
IRAS 20551-4250& HII&SB&AGN \\  
IRAS 22491-1808&HII&SB&AGN?\\ 
IRAS 19254-7245&Sey2&AGN&AGN\\ 
Mkn 231&Sey1&AGN&AGN\\ 
 
\hline 
 
\end{tabular} 
 
\end{center} 
\footnotesize\textit{{Note. $-$
Col.(1) Object name.
Col.(2) Optical Classification (\cite{vbraito-C2:Lutz1999}; \cite{vbraito-C2:Veilleux1999}). 
Col.(3) Mid Infrared classification based on ISO  spectroscopy (\cite{vbraito-C2:Genzel1998}).
Col.(4) X-ray Classification.}} 
\end{table}

\begin{figure}[!t] 
\begin{center} 
\label{vbraito-C2_fig:fig3} 
\epsfig{file=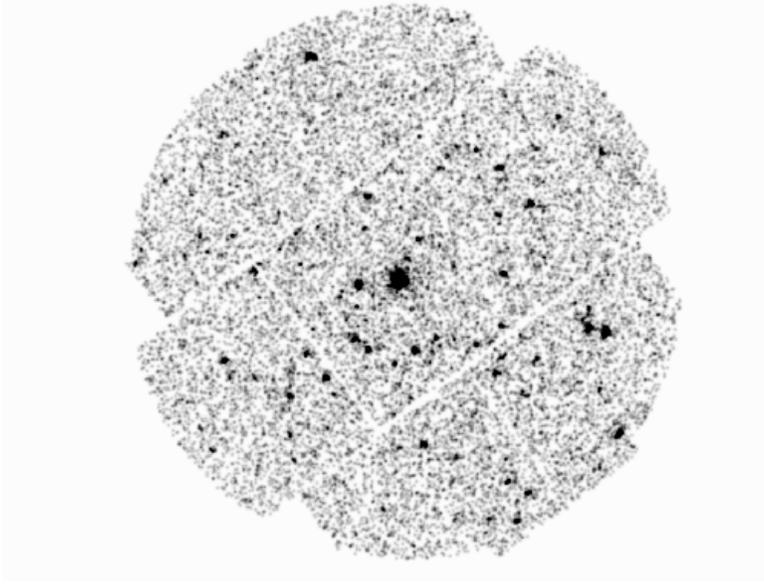, width=11cm} 
\caption{\small\textit{MOS2 ($0.2-10$ keV) image of Mkn 231 field.}} 
\label{vbraito-C2_fig:fig3} 
\end{center} 
 
\end{figure} 
 
\section{CONCLUSION} 
   
All the 5 ULIRGs observed so far have been detected in X-rays by XMM-Newton.
Although two of these sources (IRAS 20100-4156 and IRAS 22491-1808) are extremely 
faint, the presence of AGN activity cannot be ruled out by the data. 
For the remaining three sources, a clear evidence of AGN activity is found.  
In particular, for IRAS 19254-7254 we were able to detect the Iron K$\alpha$ line with an  
equivalent width $\sim 2$ keV, suggestive of a Compton-thick AGN.  
 
What is emerging from our preliminary analysis is that AGN activity seems to be present  
in the majority of the ULIRGs observed so far.
However, even if all the objects observed so far are ``Compton thick",
their hard X-ray luminosities appear to be lower than
$\sim10^{44}$ erg s$^{-1}$ even after correction for photoelectric absorption;
this would indicate that these objects are not typically type-2 QSO, but more moderate
buried AGNs.
In any case, the intermediate nature of these sources is confirmed by our XMM-Newton
survey.
\begin{acknowledgements} 
This work received financial support from ASI (I/R/037/01) under the project
``Cosmologia Osservativa con XMM-Newton" and support from the Italian Ministry of University
and Scientific and Technological Research (MURST) through grants Cofin $00-02-004$.
PS acknowledges partial financial support by the Italian {\it Consorzio 
Nazionale per l'Astronomia e l'Astrofisica} (CNAA). 
     
\end{acknowledgements}

\end{document}